\documentclass[aps,twocolumn,secnumarabic,longbibliography]{revtex4-1}

\usepackage{graphicx}
\usepackage{amsmath}
\usepackage{amssymb}

\usepackage{stackengine}
\usepackage{lineno}

\begin{document}
\title{Thermodynamic limits of sperm swimming precision.}

\author{C. Maggi}
\email[]{claudio.maggi@roma1.infn.it}
\affiliation{NANOTEC-CNR, Institute of Nanotechnology, Soft and Living Matter Laboratory, Roma, Italy}
\affiliation{Dipartimento di Fisica, Sapienza Universit\`a di Roma, Piazzale A. Moro 2, I-00185, Rome, Italy.}
\author{B. Nath}
\affiliation{Dipartimento di Fisica, Sapienza Universit\`a di Roma, Piazzale A. Moro 2, I-00185, Rome, Italy}
\affiliation{ISC-CNR, Institute for Complex Systems, Piazzale A. Moro 2, I-00185 Rome, Italy}
\affiliation{Mechanical Engineering Department, National Institute of Technology Silchar, 788010, Assam, India.}
\author{F. Saglimbeni}
\author{V. Carmona Sosa}
\author{R. Di Leonardo}
\affiliation{NANOTEC-CNR, Institute of Nanotechnology, Soft and Living Matter Laboratory, Roma, Italy}
\affiliation{Dipartimento di Fisica, Sapienza Universit\`a di Roma, Piazzale A. Moro 2, I-00185, Rome, Italy.}
\author{A. Puglisi}
\affiliation{Dipartimento di Fisica, Sapienza Universit\`a di Roma, Piazzale A. Moro 2, I-00185, Rome, Italy}
\affiliation{ISC-CNR, Institute for Complex Systems, Piazzale A. Moro 2, I-00185 Rome, Italy.}

\begin{abstract}
Sperm swimming is crucial to fertilise the egg, in nature and in
assisted reproductive technologies. Modelling the sperm dynamics
involves elasticity, hydrodynamics, internal active forces, and
out-of-equilibrium noise. Here we give experimental evidence in
favour of the relevance of energy dissipation for sperm beating
fluctuations.  For each motile cell, we reconstruct the time-evolution
of the two main tail’s spatial modes, which together trace a noisy
limit cycle characterised by a maximum level of precision
$p_{max}$. Our results indicate $p_{max} \sim 10^2 s^{-1}$, remarkably
close to the estimated precision of a dynein molecular motor actuating
the flagellum, which is bounded by its energy dissipation rate
according to the Thermodynamic Uncertainty Relation. Further
experiments under oxygen deprivation show that $p_{max}$ decays with
energy consumption, as it occurs for a single molecular motor. Both
observations are explained by conjecturing a high level of
coordination among the conformational changes of dynein motors.  This
conjecture is supported by a theoretical model for the beating of an
ideal flagellum actuated by a collection of motors, including a
motor-motor nearest neighbour coupling of strength $K$: when $K$ is
small the precision of a large flagellum is much higher than the
single motor one. On the contrary, when $K$ is large the two become
comparable. Based upon our strong motor coupling conjecture, old and new 
data coming from different kinds of flagella can be collapsed together on a simple master curve.
   \end{abstract}

\maketitle


\section{Swimming with noise}

Sperm motility plays a crucial role in sexual reproduction and also serves as a prototype for understanding the physics of
microswimmers~\cite{lauga2009hydrodynamics,elgeti2015physics}. Its investigation is fundamental to develop
new technologies, for instance to improve fertility diagnostics and assisted reproduction techniques~\cite{fauci2006biofluidmechanics}. It can also positively influence the field of artificial
microswimmers and of microfluidic devices~\cite{tsang2020roads}. 

A sperm cell is  composed by a large head (spatulate-shaped for bull sperms as those considered here) and a thin
whip-like tail called flagellum, whose oscillatory movement sustains a
travelling wave from head to  tail~\cite{lindemann2016functional}.  In the last decades, physics has investigated the sperm swimming problem,
how it originates from flagellar beating coupled to the fluid dynamics
and to the many possible boundary conditions~\cite{machin1958wave,elgeti2010hydrodynamics,gaffney2011mammalian}. Different swimming modes have been identified, including
planar beating near flat (e.g. air-liquid or liquid-substrate)
surfaces, beating with precession when the head is anchored to a
point, circular trajectories on a plane, 3d helical in the bulk, etc.~\cite{crenshaw1989kinematics}.

In modelling, minimal ingredients for
swimming of semi-flexible filaments, are
an anisotropic Stokes drag and a single travelling wave, e.g. (for
small deviations $y(t,s)$ from the straight rod shape at time $t$ and
arclength $s \sim x$) $y(t,x)=A\cos(kx-\omega t)$ which guarantees
irreversibility of the shape cycle i.e. $y(t,x) \neq y(T-t,x)$ where
$T$ is the cycle period, necessary to swim at low Reynolds numbers~\cite{gray1955propulsion,purcell1977life}.
An important element is noise, that is deviations from the {\em average} flagellum beating dynamics, which has been previously considered in modelling~\cite{julicher1997spontaneous,friedrich2008stochastic,friedrich2009steering,elgeti2010hydrodynamics}
and in experiments, with Chlamydomonas~\cite{polin2009chlamydomonas,goldstein2009noise,goldstein2011emergence,wan2014rhythmicity,quaranta2015hydrodynamics} and with sperms~\cite{ma2014active}.  In particular such experimental works have estimated through different methods the quality factor of the phase noise in the beating cycle, a parameter which is strictly connected to the precision studied here, as discussed later.  Flagellar fluctuations have been observed to influence self-propulsion~\cite{klindt2015flagellar} and  synchronization of adjacent filaments~\cite{goldstein2009noise,goldstein2011emergence,solovev2022synchronization}. 

In the present study we show how energy dissipation, an intrinsic quantity for motors at all scales, affects noise in sperm beating, rationalising the problem under the framework  of Thermodynamic Uncertainty  Relations (TURs)~\cite{barato2015thermodynamic,gingrich2016dissipation,horowitz2020thermodynamic} (see Appendix D for a summary of the simplest working principle behind TURs). 
Remarkably, the connection between power consumption and macroscopic fluctuations leads us to put forward a hypothesis  about the collective dynamics of the molecular motors actuating the flagellum. 

The sperm axoneme hosts an array of dynein molecules for a total of $N \sim 10^5$  motor domains~\cite{lindemann2003structural,chen2015atp,gilpin2020multiscale}. Each motor converts available ATP molecules into power
strokes inducing local bending of the axoneme. Deviations from the average biochemical  cycle of a molecular motor occur
mainly because of  fluctuating times of
residence in the different chemical states~\cite{bustamante2001physics}. 
Less understood is the mechanism of
coordination of the $N$ motors necessary to generate the tail's travelling
wave: a widely accepted fact is the presence of some feedback
mechanism inducing activation and de-activation of the motors based
upon the local bending state~\cite{brokaw2009thinking}. The hypothesis that a dynein operates independently of its neighbors is questioned by the observation - in micrographs by scanning electron microscopy, etc. - of non-random grouping of dynein states and by the evidence that  interactions between
adjacent dyneins may be inevitable because of the size of dynein arms~\cite{brokaw2002computer,burgess1995rigor,goodenough1982substructure}. Our experimental observations about the high amplitude of the noise affecting flagellum beating (comparable to that of a dynein motor) and about the decay of flagellum precision with energy consumption (similar to what happens for a single motor), contribute together to conjecture a strong coupling between the dynamics of adjacent  motors proteins. A schematic model for axonemal oscillations under the effect of noisy motor dynamics corroborates our hypothesis.

\section{Precision of  a Brownian motor}

\begin{figure*}
    \linespread{1.3}\selectfont
    \includegraphics[width=\linewidth]{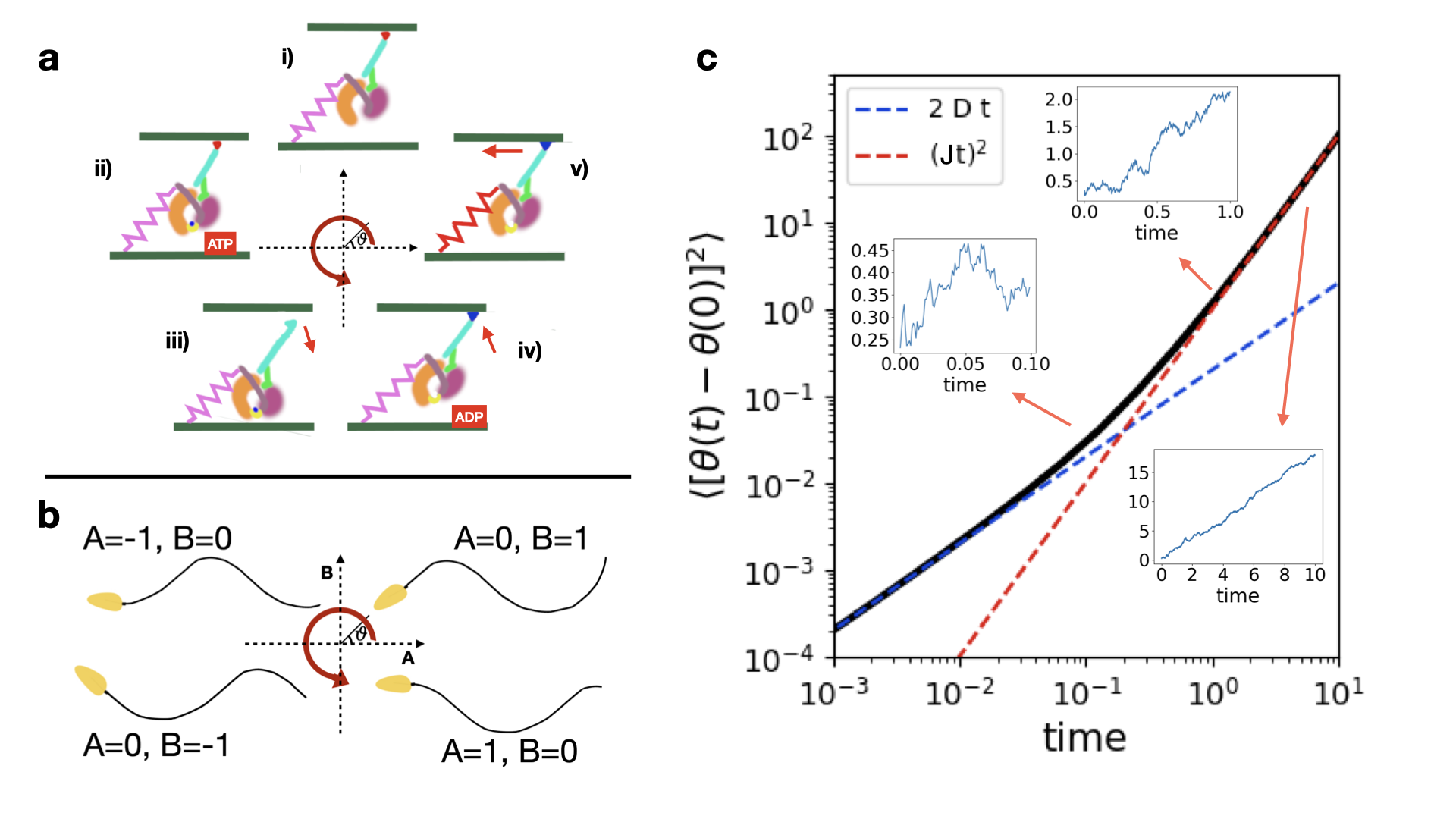} 
    \hrulefill
    \includegraphics[width=\linewidth]{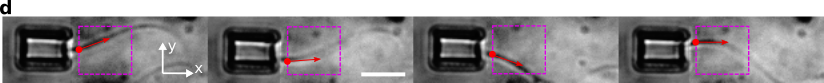} 
    \caption{\label{fig1}Precision of stochastic clocks. Examples
      - in different systems - of the coarse-grained coordinate
      $\theta(t)$ useful to define the precision $p$ of Brownian
      clocks.  {\bf a}, Sketch of the supposed chemical cycle of the
      dynein ATP-ase, anti-clockwise starting from the top: i)
      ``rest'' (apo) state, ii) ATP-binding; iii) detachment of the
      stalk; iv) ATP $\to$ ADP+P reaction with re-joining of the stalk
      to the upper substrate, in a forward position; v) stroke of the
      linker (red spring) with consequent dragging of the upper
      substrate. {\bf b}, Sketch of the elasto-mechanical cycle of a
      sperm cell, the tail shape is approximated by a curve close to
      $A(t)\cos(kx) + B(t)\sin(kx)$. {\bf c}, Mean squared
      displacement from a trajectory generated by the numerical
      integration of equation $\dot \theta=J+\sqrt{2D}\eta(t)$ with
      $\eta(t)$ white noise with unitary amplitude, with $J=1$ and
      $D=0.1$. The insets show the trajectory $\theta(t)$ with a
      different time-range, i.e. c1) $t\in[0,0.1]$, c2) $t\in[0,1]$, c3) $t\in[0,10]$. It is evident how the mean current
      $\overline{\dot\theta}=J$ can be appreciated only at times much larger than $1/p=D/J^2$. {\bf d}, successive
      snapshots (taken at a distance $0.04$ seconds with an optical
      microscope, see Appendix A) of the tail dynamical sequence in a caged sperm
      experiment: the purple box represents the region of interest,
      the red dot and red arrow identify the parameters $a,b$
      (y-position and slope of the tail close to the origin,
      respectively) which approximate $A,B$ in Eq.~\eqref{apdyn}, here corresponding to $(A,B) \sim (0,1) \to (-1,0) \to (0,-1) \to (1,0)$. The white arrows indicate $\hat{x}$ and $\hat{y}$ directions  ($\hat{z}$ direction is perpendicular to both). The white scalebar represents $10 \mu m$.}
\end{figure*}

We first discuss how to measure ``precision'', an observable which has recently attracted a profound
interest in non-equilibrium
statistical physics, see Fig.~\ref{fig1}. For our purpose it
is sufficient to consider a system where an angular observable $\theta(t)$ represents the system’s configuration (see Fig.~\ref{fig1}a and b). We expect $\theta(t)$ to perform an
irreversible stochastic stationary dynamics with average drift $\langle \theta(t) -\theta(0) \rangle = Jt$ and relative dispersion
$\langle [\theta(t)-\theta(0) - Jt]^2 \rangle \sim 2Dt$ for large
$t$. We are interested in the precision
rate defined as
\begin{equation}
  p=\frac{J^2}{D}.
\end{equation}
The observable $p$ can be understood as the inverse of the typical
time $t^*=1/p$ separating the diffusive regime $t \ll t^*$  ($Dt \gg (Jt)^2$) from the ballistic regime $t \gg t^*$  (see Fig.~\ref{fig1}c and its insets).

The quantity $p$ has been demonstrated - through the so-called
Thermodynamic Uncertainty Relation
(TUR)~\cite{barato2015thermodynamic,gingrich2016dissipation,horowitz2020thermodynamic}
to be bounded from above by the entropy production rate, or in
practical terms (for the purpose of steady isothermal molecular
motors) the motor's energy consumption rate $\dot{W}$ in thermal
units:
\begin{equation} \label{tur}
p \le p_{TUR}=\frac{\dot{W}}{k_B T}.
\end{equation}
The ratio   $\mathcal{Q}=p/p_{TUR} \le 1$
can be considered as a motor's figure of merit. Estimates of
$p$ through Markovian models informed by experimental data~\cite{hwang2018energetic} suggest
that several molecular motors work not far from their optimum, or at least close to its order of magnitude ($\mathcal{Q}  \ge 0.1$).   In the following we present a method to estimate $p$ and we apply it to experiments with bulls' sperm cells (see Fig.~\ref{fig1}b and d). Notwithstanding its physical relevance, the quantity $p$ has not been discussed for
microswimmers, even if its estimate can be deduced from other variables in previous works.
Quantities which are strictly related to $p$ are the dissipation time~\cite{falasco2020dissipation} and the quality factor $q=J/(2D)=p/(2J)$, which has been measured within a similar approach for Chlamydomonas flagella in~\cite{polin2009chlamydomonas,goldstein2009noise,goldstein2011emergence,wan2014rhythmicity,quaranta2015hydrodynamics} and for bull sperms in~\cite{ma2014active}, although  never compared to energy dissipation and or discussed within the framework of TURs. 

\section{Precision of sperm beating  }

We adopt a coarse-graining protocol that reduces the  space
of possible shapes of the flagellum 
into two coordinates, the minimum for the existence of irreversible limit cycles. We improve the quality of image
tracking and disentangle the simplest mode of sperm movement - that is
the planar one -  with the following technique: each observed sperm has its
head trapped in a microcage printed by 2-photon microlitography, see
Fig.~~\ref{fig1}d and Appendix A.  The cell cannot spin and
the flagellum beats on the $\hat{xy}$ plane. While the most common swimming strategy of sperm cells is helical~\cite{rikmenspoel1960cinematographic,crenshaw1989kinematics}, planar movement is 
typically observed close to a surface and can lead to circular paths~\cite{woolley2003motility,nosrati2015two,saggiorato2017human}, here prevented by the cage.  The sperm’s center of mass has a very limited dynamics in
the $\hat{xz}$ plane but can oscillate along the transverse $\hat{y}$
direction; the body, entirely free, performs planar tail
beating  which pushes the body into the cage making the escape probability negligible. As a direct consequence of tail beating, the head is also
observed to oscillate: our main results are obtained by
tail tracking, while in the Supplementary Information we
confirm our conclusions by tracking the head, see~\cite{suppl} and its Fig. S1.

\begin{figure*}
    \centering
           \includegraphics[width=0.31\linewidth]{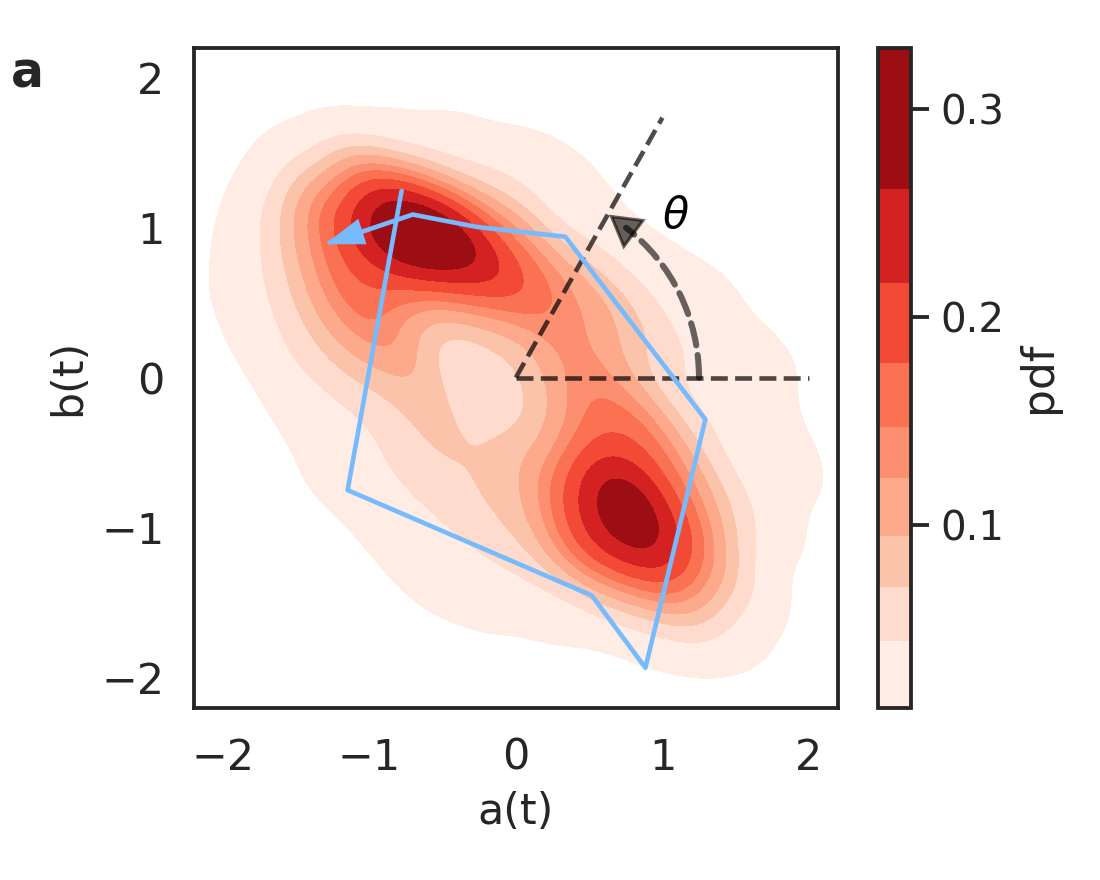}  
           \stackinset{l}{37.5pt}{t}{11pt}{\includegraphics[width=1.8cm]{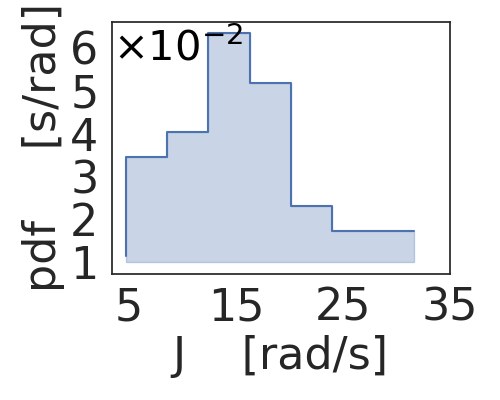}}{\includegraphics[width=0.31\linewidth]{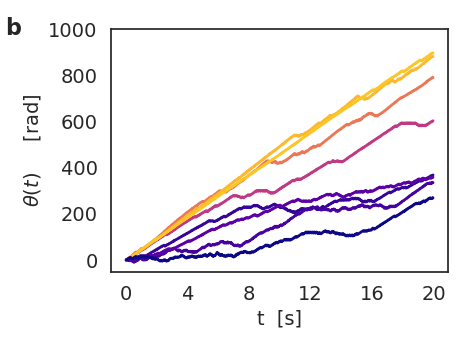}}
           \stackinset{r}{17pt}{b}{29pt}{\includegraphics[width=2cm]{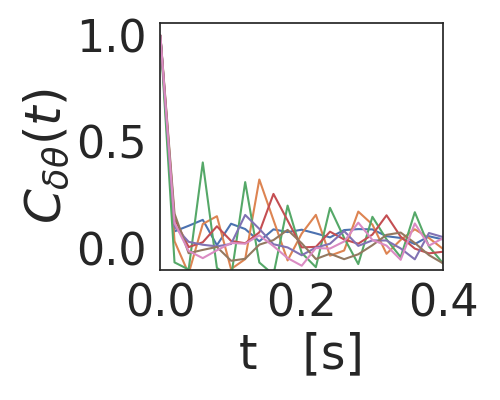}}{\includegraphics[width=0.31\linewidth]{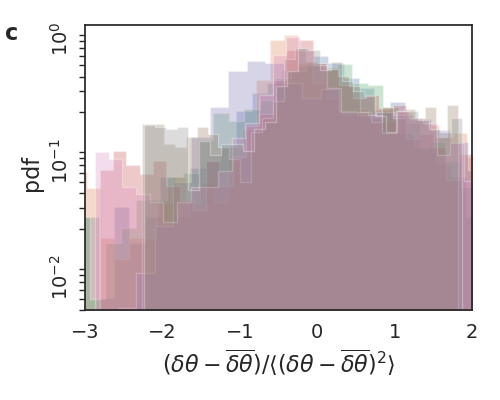}}\\
           \stackinset{l}{33pt}{t}{9pt}{\includegraphics[width=2cm]{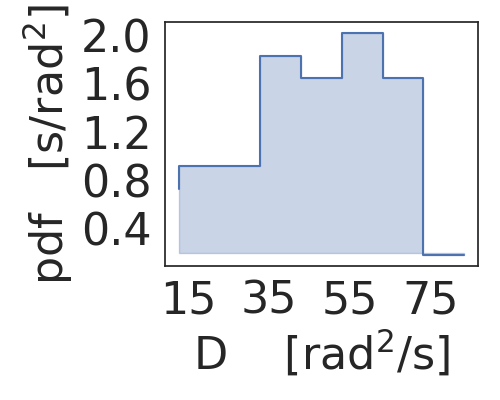}}{\includegraphics[width=0.31\linewidth]{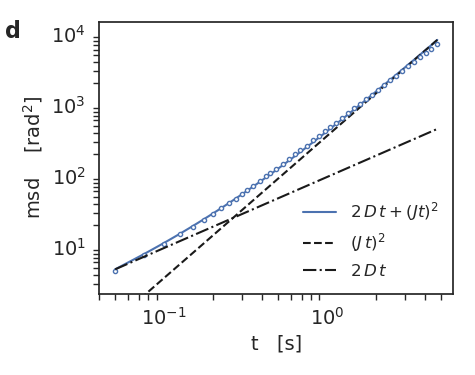}}
           \includegraphics[width=0.31\linewidth]{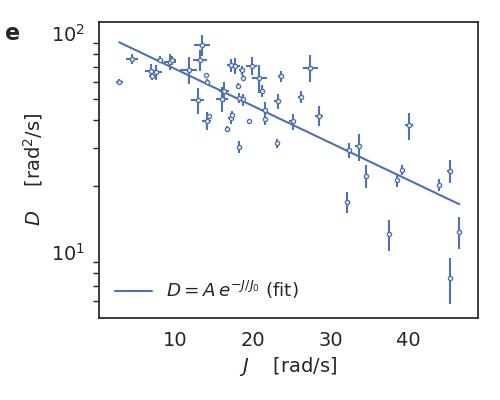}
           \stackinset{l}{36pt}{b}{27pt}{\includegraphics[width=1.95cm]{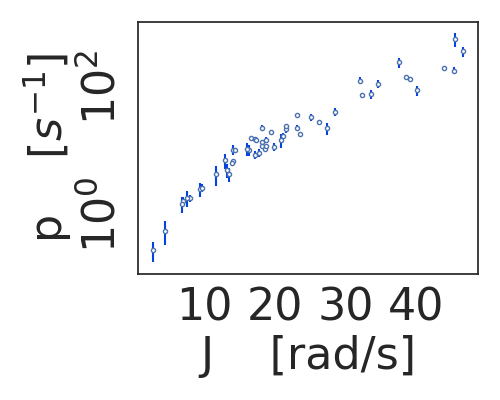}}{\includegraphics[width=0.31\linewidth]{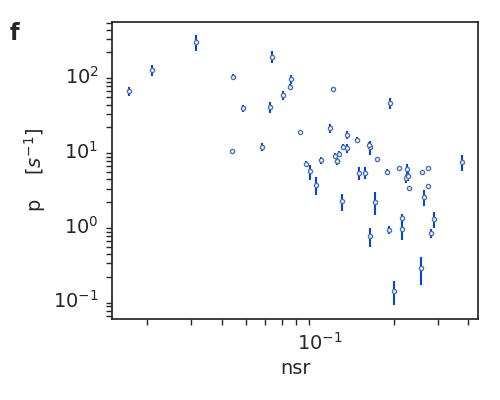}}
\caption{\label{fig2}Extracting precision from tail tracking. {\bf
    a}, Histogram of the positions in the plane $a(t),b(t)$ for the $12$ sperms
  with precision $p>20 s^{-1}$. The light-blue curve traces the $a(t),b(t)$ path for a particular sperm in $0.2$ seconds ($10$ frames). {\bf b}, Integrated phase-space current $\theta(t)$ for a few
  observed sperms. The inset shows the probability density (pdf, over all observed sperms) of the average current $J$. {\bf c}, Pdf
  (over $20$ seconds acquisition), for a few sperms, of $\delta
  \theta=\theta(t+dt)-\theta(t)$ (shifted by the mean and scaled by
  the standard deviation), where $dt=0.02$ seconds. The inset shows a few autocorrelations $C(t)=\langle \tilde{\delta\theta}(t)\tilde{\delta\theta}(0)\rangle$ where $\tilde{\delta\theta}=\delta\theta-\langle \delta\theta\rangle$. {\bf d}, Mean
  squared displacement of $\theta(t)$ for a given sperm, and its fit
  according to the model $msd(t)=2Dt+(Jt)^2$. The inset shows the
  pdf of diffusivity $D$ over all observed
  sperms. {\bf e}, Diffusivity $D$ versus average current $J$ together with decaying exponential fit $D \sim 10^2*e^{-J/25} s^{-1}$. {\bf f}, Precision
  $p$ versus the noise-to-signal ratio (nsr) computed from the signals
  $a(t)$ and $b(t)$. The inset shows the $p$ versus the current $J$. Error bars for $J$ and $D$ (in plots e and f) denote $3$ times the standard deviation of $J$ and $D$ estimated when fitting the msd by chi-square optimization. Data come from the observation of $54$ different sperm cells, if not differently specified. }
\end{figure*}

Referring to Fig.~\ref{fig1}d, our region of
interest (ROI) tracks less than half of the observed beating wavelength. After image processing (see Appendix B) each tail's image
is fitted through a second order polynomial $y(x,t)=a(t)+b(t)x+c(t)x^2$ (see also Movie  in  the Supplementary Information~\cite{suppl}). Under
the assumption that the ROI contains less than half wavelength (and
therefore $y(x,t)$ has at most one extremal point), $a(t)$ and $b(t)$
are - but for multiplicative constants - fair approximations of the
coefficients $A(t)$ and $B(t)$, respectively, of a mode expansion
\begin{multline} \label{apdyn}
y(x,t) \approx A(t)\cos(kx) + B(t)\sin(kx) \\ \sim_{x \to 0} A(t)+kB(t) x + \mathcal{O}(x^2)
  \end{multline}
Such a kind of shape approximation and the consequent coarse-graining of
the planar flagellum dynamics into two main coordinates $A,B$ has been
used for bull sperms~\cite{ma2014active}, with Chlamydomonas flagella~\cite{polin2009chlamydomonas,goldstein2009noise,goldstein2011emergence,wan2014rhythmicity,quaranta2015hydrodynamics} and with human
sperms~\cite{saggiorato2017human}. A similar approach to  the breakdown of detailed  balance in flagella has been adopted
in experiments with Chlamydomonas~\cite{battle2016broken}, with filaments in actin-myosin networks~\cite{gladrow2017nonequilibrium}, with {\it C. elegans}
worms~\cite{stephens2008dimensionality}. A general perspective
about this strategy is discussed in a recent review~\cite{gnesotto2018broken}. In experiments of this kind, however, precision and TUR are rarely considered~\cite{li2019quantifying,roldan2021quantifying}.

For the purpose of estimating $p$ for each cell, we first apply a
filter to the $a(t),b(t)$ time series in order to remove low-frequency
drifts, including average, and normalise the data to have a standard
deviation of $1$. We observe that the two
coordinates exhibit almost harmonic oscillations at a similar frequency
$f \sim 6 - 8 Hz$ (see  spectra in Appendix B, Fig.~\ref{figM3}b,c) but
with a phase delay $\Delta(t)$ that fluctuates around a steady non-zero
value (Fig.~\ref{figM3}a). This delay allows us to reconstruct the angle
$\theta(t)$ in the plane $a(t),b(t)$, see Fig.~\ref{fig2}a, and finally measure
the average phase-space current $J=\langle \theta(t)\rangle/t$, see Fig.~\ref{fig2}b.  Apart from
a few noise dominated cells where $J$ is small and negative we find $J>0$, as expected from the geometrical
interpretation of $a(t)$ and $b(t)$ in terms of the main modes of the
tail’s shape. A negative $J$ would correspond
to a  waveform travelling in a  direction that is incompatible with
forward swimming.  We stress that the cumulative phase-space angle
$\theta(t)$ is proportional to the number of performed cycles of the
sperm’s tail shape dynamics. The average current $|J|$ is related to the beating frequency $f$ in a
subtle way: in fact, the growth of $\theta(t)$ is influenced not only
by the oscillation of $a(t)$ and $b(t)$ but also by the sign of their
phase delay $\Delta$. Failures to guarantee a constant sign
of $\Delta$ imply {\em ineffective beatings}, i.e. uncoordinated
oscillations which do not contribute to the growth of $\theta(t)$, leading to $|J| \le J_{max}=2\pi f$.

Fluctuations of the rotation speed $\dot\theta(t)-J$ are well visible in our experiment, see Figs.~\ref{fig2}b and c, and represent departures from the average shape cycle
$\theta_{ideal}(t)=Jt$. They are in part due to real dynamical noise
(“stochastic deviations”) and in part  to the fact that the real
shape dynamics is slightly different from the approximated one
(“deterministic deviations”). Since deterministic
deviations are periodic and each experiment includes hundreds of
beating periods, their contribution to the diffusivity $D$
can be safely neglected for our purpose. The main origin of stochastic deviations is non-equilibrium
fluctuations, acting both on the  fluid surrounding the flagellum and on the working cycle of the thousands
of molecular motors actuating the flagellum. At low
Reynolds numbers the first effect is negligible (see Appendix B.2).

We empirically find a good fitting model for the mean squared displacement (msd) $\langle
[\theta(t+\tau)-\theta(t)]^2 \rangle \sim J \tau + 2D\tau^2$ (averaged
over $t$ along each whole experiment), see Fig.~\ref{fig2}d for an example.  Alternative ways to estimate the diffusivity are discussed in~\cite{ma2014active}, we have considered them for our experiment, finding a substantial agreement, see the Supplementary Information~\cite{suppl} with Fig.~S2. In Fig.~\ref{fig2}e we show the relation between diffusivity $D$ and average
current $J$ displaying an average decay but with wide population variability. In Fig.~\ref{fig2}f we plot the measured values of $p$ in a large
set of experiments, as function of the noise-to-signal ratio (defined as the ratio between the peak of the  spectrum and its average at frequencies higher than the oscillation frequency, see Fig.~\ref{figM3} in the Appendix B), and - in the inset - versus
$J$. Our first main conclusion is that $p$ takes values in the
approximate range $0- p_{max}$ with $p_{max} \approx 10^2
s^{-1}$. Moreover we see that it roughly decreases with the
noise-to-signal ratio and it roughly increases with $J$. A visual inspection of the extremal cases, i.e. those close to $0$ and those close to
$p_{max}$, confirm that they correspond to  chaotic
motion and to almost regular  periodic motion respectively.

\section{Sperm precision is much smaller than the TUR bound}

Direct empirical estimates of the energy consumption - through
respiration and glycolysis - for various types of 
sperms, gave figures in the range of $10^7 \div 10^8$ $k_B
T/s$, see Appendix B~\cite{brokaw1967adenosine,rikmenspoel1969energy,chen2015atp}. 
Theoretical estimates for the power produced by a
micro-swimmers are given by the Taylor formula, here adapted for bull sperms~\cite{rikmenspoel1969energy}:
\begin{equation} \label{taylor}
\dot{W}_{prod} \approx \pi^3 \eta L f^2 \beta^2,
\end{equation}
where $\beta$ is the tail beating amplitude, $L$ is the tail length and $\eta$ the host fluid viscosity (we have assumed, in the original Taylor’s formula,  the cross section of the flagellum to be $\sim 0.2 \mu m$ and the tail wavelength $\sim 35 \mu m$).
We set $\beta=5\mu m$ (only weakly varying with
external conditions in our range, see~\cite{rikmenspoel1984movements}), $L=60 \mu m$, $\eta=10^{-3} \textrm{Pa} \cdot s$. 
The accepted order of magnitude of sperm’s efficiency
is $\sim 10-25\%$~\cite{brokaw1966effects,brokaw1975effects,nicastro2006molecular,carvalho2011evolution,klindt2016load,pellicciotta2020entrainment}, giving values which are compatible with experimental estimates  $\dot W \approx 10^8 k_B T/s$ when $f=20 Hz$ at $37^\circ C$. In our experiments at room temperature $\sim 20^\circ C$  the typical beating frequency is $6-7 Hz$, leading to $\dot W \approx 10^7 k_B T/s$, and  therefore $p_{TUR}=10^{7} s^{-1}$. It is clear that all these figures rest in a much narrower range if the normalised consumption rate is considered $\dot W/f^2 \approx 2-5 \cdot 10^6 k_B T s$. In conclusion the bound Eq.~\eqref{tur} largely overestimates our measured maximum precision, with $\mathcal{Q}_{macro}
=p_{max}/p_{TUR} \sim 10^{-5}$. In the following we propose an interpretation of this result.

An intriguing observation concerns the maximum precision,
computed from empirical data-informed models, of the dynein molecular
motors $p_d$~\cite{hwang2018energetic} which is  close to
the maximum values we have measured for the whole flagellum $p_d \sim
p_{max}$~\footnote{We remark that in this paper~\cite{hwang2018energetic}
  cytoplasmic dynein is considered, which is known to be structurally
  similar to the axonemal one, with also a few distinct
  features~\cite{kato2014structure}. }.  Our interpretation of the
similarity between those two figures is the following. Let us denote
with $\theta_i(t)$ the integrated current - in the space of motor
configurations - in the time $t$ for the $i$-th dynein motor. In both the systems
(sperm’s flagellum and dynein) the current integrated in time counts
the cumulative number of performed cycles in the
 configuration space. We conjecture  that - in a given
amount of time - the number of cycles in the configuration space of
the sperm’s flagellum is proportional to the number of cycles in the
motor configuration space of {\em any} molecular motor in that
flagellum, i.e. $\theta(t) \approx C \theta_i(t)$, $\forall i
\in[1,N]$, with possible $i$-dependent corrections which 
rapidly vanish with $t$. The result of this conjecture is that the
precision of variable $\theta(t)$ is close to the precision of
variables $\theta_i(t)$ for any $i$. The biological meaning of our
conjecture is that a long-range coordination among molecular motors inside the
flagellum, quite an accepted fact in the
literature~\cite{brokaw2009thinking}, affects also  fluctuations. 
In order to make our conjecture more robust, we proceed along two different roads: a  new theoretical model and a second experiment. 

\begin{figure}
    \centering
     \includegraphics[width=\linewidth]{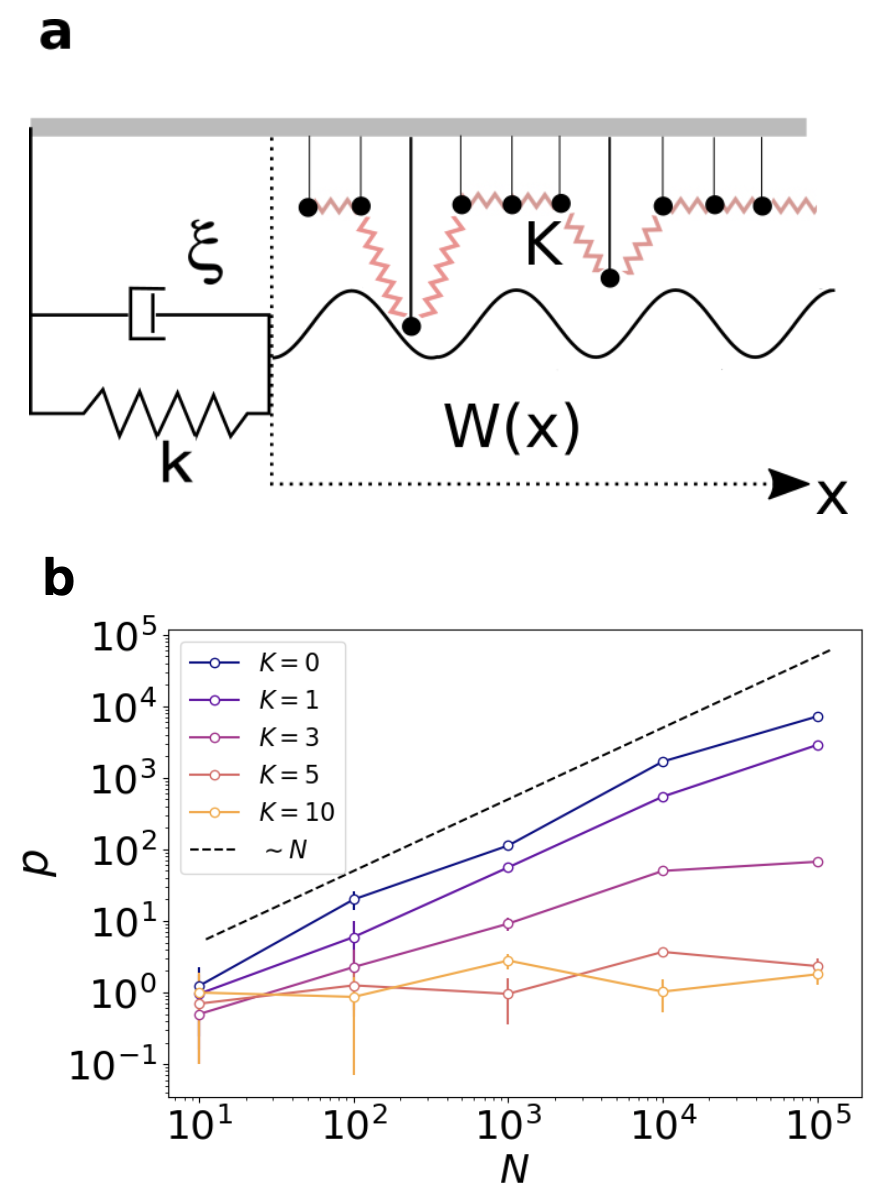}  
    \caption{\label{fig4}Theoretical model. {\bf a}, sketch of the model; {\bf b}, precision $p$ versus number of motors $N$ for different choices of the coupling  parameter $K$, showing  the $O(N)$ scaling for uncoupled motors ($K=0$) and the $O(1)$ scaling for large $K$.  In all the simulations we have used $\alpha=\eta=0.5$, $k/(\xi \Omega)=10$, $\alpha N U_0/(\Omega \ell^2 \xi)=0.6$. The error bars are obtained to error propagation based upon the error in the measurement of $D$. Such error is the estimated standard deviations of $D$ in the non-linear least squares fit  of the exponential decay of the phase correlation (see Appendix C). }
\end{figure}

\begin{figure*}  
    \centering
   \includegraphics[width=\linewidth]{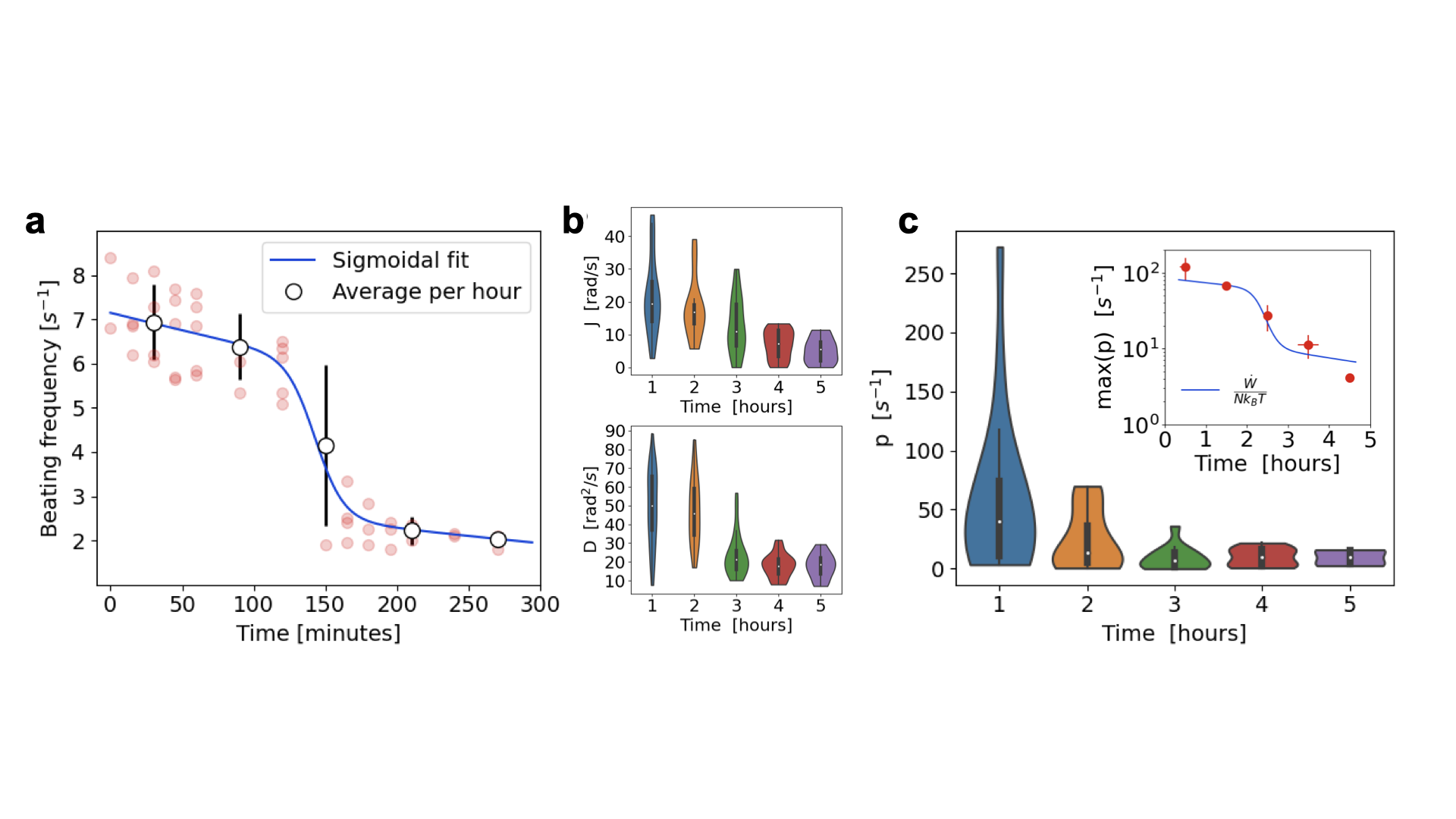}  
\caption{\label{fig3}Sperm precision and thermodynamics.
  Observations for caged sperms in experiments within a sealed
  chamber. {\bf a}, Decay of beating frequency (as measured from the
  spectrum peak of $a(t)$ and $b(t)$), with time. The averages values
  of the frequencies at each hour of experiment are also marked as
  white circles, together with a sigmoid-like fit $f(t) = c_1 e^{-c_2
    t}s(t)+c_3 e^{-c_4 t}[1-s(t)]$ with sigmoid
  $s(t)=[1+e^{(t-t_0)/\tau}]^{-1}$ and best fit values $t_0=143$,
  $\tau=10.2$, $c_1=7.1$, $c_2=1.1 \times 10^{-3}$, $c_3=3.5$, $c_4=2 \times 10^{-3}$. {\bf b,}
  reduction of average current and diffusivity along time, shown as
  violin plots each from a hour bin (e.g. $1$ means all observation
  done in the $1$st hour, etc.). {\bf c,} decay of precision with
  time, again in the form of a violin plot. The inset shows the decay
  of $p_{max}$ (estimated as the average of the top $25\%$ population)
  together with the normalised (by $N=10^5$) power consumption
  extracted by the Taylor formula, eq.~\eqref{taylor} using the
  frequency decay fit shown in Fig. 3a, with amplitude fixed at the
  average observed value $5.2 \mu m$. Error bars represent standard
  deviations normalised by square root of the number of data in the
  aforementioned percentile. The sizes of samples in the plots of
  frames b and c are: 41 sperms in the 1st hour, 24 in the 2nd hour,
  34 in the 3rd hour, 25 in the 4th hour and 16 in the 5th hour.
   }
\end{figure*}

\section{A model with strongly coupled motors}

A first clue in support of our conjecture comes from the numerical
analysis of a theoretical model for the motor-actuated flagellar
dynamics, extensively studied
in~\cite{julicher1997spontaneous,guerin2011dynamical,guerin2011bidirectional},
modified here through the introduction of a coupling term between
adjacent motors. The model is depicted in Fig.~\ref{fig4}a and is
described in details in the Appendix C. It consists of a filament
with $N$ motors. Each motor acts on the filament through an
interaction potential, and performs a stochastic attachment/detachment
dynamics which breaks detailed balance as if consuming ATP. The motor
position $X$ oscillates under the joint effect of the forces of the
attached motors and of an external elastic force $\xi\dot
X(t)=-\partial_X \sum_i s_i U[x_i-X(t)]-\kappa X$ with $s_i \in
\{0,1\}$ representing the detached-attached status of $i$-th motor,
the motor-filament potential $U(x)=U_0[1-\cos(2\pi x/\ell)]$,
viscosity $\xi$ and elastic constant of the external spring $\kappa$.
The elastic force here could represent the effect of the cage but in
previous studies was introduced just to simplify the mathematics of
the problem, it is not crucial for the model's
phenomenology~\cite{guerin2011dynamical}. The variables $s_i$ jump
from $0$ to $1$ and back according to a Poisson process. In the
original model the probability rates of such a process depended only
upon the local motor-filament potential, so that the fluctuations of
the jump dynamics of each motor was independent from nearby motors:
for this reason the amplitude of the macroscopic noise was observed to
decrease with $N$~\cite{ma2014active}. Here we employ a binding
potential $K(s_i-s_{i+1})^2$ that correlates the states $s_i$ and
$s_{i+1}$ of adjacent motors. Increasing $K$ (from the case $K=0$
which corresponds to the original version) drastically changes the
behavior of the model, in particular resulting in a much stronger
macroscopic noise, i.e. a largely faster decay of the phase
correlation, see Fig.~\ref{figM2} in the Appendix C. 

 In Fig.~\ref{fig4}b
we draw our main new conclusion, measuring the precision $p=\omega^2/D$
where $\omega$ is the average oscillation frequency and $D$ the
diffusivity deduced from the decay of phase
correlation~\cite{ma2014active}. When $K=0$ the precision grows linearly with $N$, in agreement with what already observed in~\cite{ma2014active} and with the reasonable argument that the $N$ random independent fluctuations of the motor phases contribute with a variance $1/N$ to the fluctuations of the macroscopic phase. However, the noise reduction
due to the growth of $N$ disappears at large $K$: when $K$ increases
the size-scaling of $p$ goes from $p \sim O(N)$ to $p \sim
O(1)$. This result amounts to say that the precision of the whole
flagellum becomes comparable to the precision of the single motor when
$K$ is large enough, as in the experiment. On the other side, the energy consumption (ATP
consumed per cycle and per motor) increases with $N$, mostly
independently of the coupling strength. The ratio between precision $p
\sim O(1)$ and energy consumption $\dot W \sim N$ therefore decreases
as $1/N$, in fair agreement with our experimental observations.

\section{Experiment under oxygen deprivation}

Experimentally, we reconsider the
TUR, Eq.~\eqref{tur}. For a single
dynein motor, in fact, it establishes a {\em close} upper bound: $\mathcal{Q}_{micro} =
p_d/(\dot W_{dynein}/k_B T) \gtrsim 10^{-1}$. The closeness of the bound suggests that a variation
of energy consumption must reflect into a proportional variation of
dynein’s precision, confirmed also in theoretical models~\cite{hwang2018energetic}. Therefore, if the noise of the
flagellum beating is dominated by molecular motors' noise, a reduction of energy consumption 
should reflect into a reduction of the flagellum’s $p_{max}$.

We have performed a series of experiments in oxygen deprivation, see
Fig.~\ref{fig3}. The samples were let in a sealed box for several
hours, recording activity and assessing the $p$ of all trapped cells,
every $15-30$ minutes. During the total time of the experiment ($5$
hours) we observed a clear decay in the beating frequency $f$, see
Fig.~\ref{fig3}a.  Although we cannot directly control if the
reduction of beating frequency is induced only by the reduction of
oxygen or of other nutrients, sperms clearly reduce their activity and
- as a consequence - their energy consumption.  During the experiment
we also observed a decay of both $D$ and $J$ (see Fig.~\ref{fig3}b)
and most importantly of the maximum precision $p_{max}$ by more than a
order of magnitude, see Fig.~\ref{fig3}c. Remarkably the observed
decay of $p_{max}$ is well reproduced by the decay of energy
consumption {\em normalised} by $N$, i.e. $p_{max} \approx \dot{W}/(N
k_B T)$, where $N=10^5$ is an estimate of the number of dynein motor
domains in a
flagellum~\cite{lindemann2003structural,chen2015atp,gilpin2020multiscale},
see blue solid and dashed lines in the inset of Fig.~\ref{fig3}c. We
interpret this result as an argument in favour of the conjecture that
fluctuations in the flagellum beating are dominated by fluctuations of
spatially correlated dynein motors.

\section{Generalisation to other eukaryotic flagella: a TUR-based correlation length}

\begin{figure} 
    \centering
  \includegraphics[width=\linewidth]{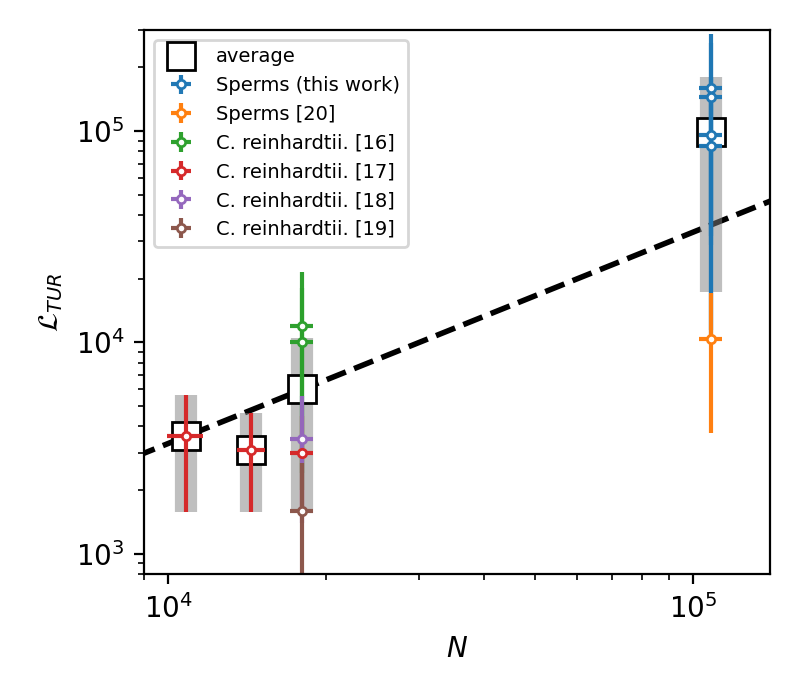}
  \caption{\label{figM1}Collapse of correlation lengths in several
      experiments. The data from different previous experiments with
    sperms and flagella of C. reinhardtii algae are summarised here by
    plotting the correlation length $\mathcal{L}_{TUR}=\frac{\dot{W}}{p k_B T} $, discussed in
    the Appendix B.3, versus the length of the
    flagella $N$ (both quantities are given as numbers of molecular
    motors). The data are also reported in
    Table~\ref{summarytable}. The light colored points allow to
    distinguish the different experiments, while the solid circles
    identify the averages at each given flagellar length. Bars
    represent statistical errors. The dashed line marks the  scaling law $\mathcal{L}_{TUR} \sim N$ which is expected under the hypothesis of strong coordination among adjacent motors in the axoneme.}
\end{figure}

We underline that, in order to extrapolate it to other systems and
more general conditions, the identification with $N$ for the ratio
between the TUR bound and the actual sperm precision should be taken
as an order of magnitude.  Here we discuss this point in closer
details.  We propose to generalise our observation to assemblies of
$N$ molecular motors in the form $p_{max} \approx
\dot{W}/(\mathcal{L}_{TUR} k_B T)$ where $\mathcal{L}_{TUR} \sim N$ is
a correlation length (measured in adimensional units, i.e. as an
estimate of the number of adjacent correlated motors). In the Appendix
B.3 section we show that such a generalisation follows by considering
a chain of molecular motors whose dynamics is correlated up to an
extension of $\sim \mathcal{L}_{TUR}$ adjacent motors, leading to a
renormalisation of the precision by a factor $\mathcal{L}_{TUR}$. In
order to corroborate our conjecture, we reconsidered several previous
results where the quality factor for fluctuations was measured in
different conditions and with different flagella (from sperms and
C. reinhardtii
algae)~\cite{goldstein2009noise,goldstein2011emergence,wan2014rhythmicity,quaranta2015hydrodynamics,ma2014active}. A
summary of our and previous results is given in
Table~\ref{summarytable}, in Appendix B.4.  Our conjecture allows to
collapse new and old data upon a master curve $\mathcal{L}_{TUR} \sim
N$, fully consistent with our hypothesis, see Fig.~\ref{figM1}.


\section{Conclusions and  outlook}

We have reported an experimental protocol to estimate the statistical
precision of sperm’s beating, which differs from previous measurements of the quality factor as it is directly related to energy  consumption, according to the   recently celebrated TURs.   The use of single-cell traps aids the
reconstruction of the dynamics of a single cell’s shape, but in future
implementations it could be replaced by a  comoving tracking
analysis directly applied upon free swimming cells. 

Our results point to the need of understanding dynamical fluctuations
of active flagella and their relation  to their bioenergetics~\cite{skinner2021improved,yang2021physical,tan2021scale}. It seems that a
recognised theoretical statement, the Thermodynamic
Uncertainty Relation, has a relevance not only for molecular
motors, but also for mesoscopic self-propelling microswimmers.  With this aim, we have reported two striking
observations: 1) the coincidence between the maximum precision of the
whole sperm cell and that of molecular motors actuating the sperm’s
flagellum, 2) the dependence of the maximum precision of the whole
sperm cell upon the reduction of energy consumption, a dependence that
one would expect only for the molecular motors. As a common explanation we conjecture that the $N \approx 10^5$ dynein motors
actuating a sperm’s tail work at a high level of coordination which also affects fluctuations: a theoretical model where adjacent motors are coupled by a binding potential is consistent with out observations. The TUR is therefore still valid for the whole sperm’s cell, but with a discrepancy between maximum
precision and energy consumption which is $\sim N$ times worse than in
the case of the single molecular motor. An interesting perspective involves studying the
same observables with other microswimmers, such as {\em E. coli}, whose flagellar motor fluctuations have been studied in the past~\cite{samuel1995fluctuation} but not their connection with the TUR.
It will also be important to understand more deeply the detailed mechanical modelling of ciliary oscillations and how fluctuations can emerge from the dynamical instabilities that underlie the axonemal beating~\cite{riedel2007molecular,sartori2016dynamic,mondal2020internal}.
Validating such models will require a comparison with the full pdf of the beating phase fluctuations (and not only its extreme values).

We conclude emphasizing that our work suggests new applications of the TUR and of the precision observable. First, the TUR let us evaluate if the observed precision is low or not, as it  gives a theoretical bound which can be reached by certain systems (for instance, some kinds of molecular motors get quite close to it). A large distance from such a bound is an observation which stimulates further investigation. Second, the precision $p$ helped us in validating the theoretical model: the scaling of $p$ with $N$ suggests the relevance of the coupling ingredient, beyond any precise calibration of the model parameters. Finally, $p$ could be made useful - in the future - in fertility studies and diagnostics, as it can enter the list of parameters measured in a spermiogram, to assess the health of human or animal sperm: for instance,  our study suggests that it is correlated to the energy consumption of the cell. Of course the usefulness of such a parameter (e.g. if it is correlated with the good performance of a cell in chemotaxis, or other kinds of migration mechanisms) must be validated with further studies.

\begin{acknowledgments}
 A. P. and B. N. acknowledges the financial support of Regione
Lazio through the Grant "Progetti Gruppi di Ricerca" N. 85-2017-15257
and from the MIUR PRIN 2017 project 201798CZLJ.
\end{acknowledgments}



\section*{Appendix A: Experimental}

\subsection*{Microfabrication}

The micro-cage features allowed to accommodate one single cell on it, in a way that the head is confined while leaving the entire tail outside. Based on sperm characteristics, the chamber is designed as a box composed by four microfabricated facets anchored to the cover glass. The height, width and depth of a single cage are 500$\,nm$, 5.5$\,\mu m$ and 11$\,\mu m$, respectively.
Microfabrication was carried out by a custom built two-photon polymerization setup \cite{Gaszton2017}. The micro-chambers were generated from SU-8 3025 photoresist (Kayaku Advanced Materials) using a 60x 1.4NA objective. After exposure, the photoresist sample was baked ramping the temperature from 65 $^{\circ}$C up to 95 $^{\circ}$C with increments of 5 $^{\circ}$C per min, then 7 min at the highest temperature. Reduction of stress between the substrate and SU-8 is achieved by gradually decreasing the temperature of the sample until reaching room temperature. Thereafter, the photoresist was developed by its standard developer solvent, followed by rinsing in a 1:1 solution of water and ethanol, and finally dried with a gentle blow of nitrogen. Strong adhesion of the micro-chambers to the carrier cover glass was ensured by three layers of Omnicoat adhesion promoter (Kayaku Advanced Materials). Laser power and scanning speed were 5 mW and 30 $\mu m s^{-1}$, respectively.

\subsection*{Sample preparation}

The experiments for measuring the main spatial modes of the sperm's tail were developed on an open sample. This sample was obtained by attaching a plastic ring surrounding the micro-chambers area using the optical adhesive NOA81 (Norland Products Inc.). For the experiments under oxygen deprivation we used hermetically sealed samples. Two fishing wires of $\sim$100$\,\mu m$ thickness and NOA81 adhesive were used as spacers between the carrier cover glass and a cover slip, generating a channel; after introducing approximately 150$\,\mu L$ of solution containing sperm cells, the sample was completely sealed by applying NOA81 adhesive in the two open sides.
\indent Bull sperms were obtained from “Agrilinea S.R.L.” (Rome), and stored in a liquid nitrogen cylinder. On the day of the experiment, a vial of sperms suspended in semen was taken and immersed in a hot water bath of 37 degrees centigrade for 10 minutes. The vial was then taken out of the bath and immediately cut open using a pair of sterilized scissors. The sperm suspension was poured out of the vial in an Eppendorf. A micropipette was then used to suck out 150$\,\mu L$ of the sperm suspension from the Eppendorf and insert the fluid into the microchannel, ensuring proper filling inside the structures. The sperm movement was recorded at environmental temperature, $\sim 20^\circ C$, by using a digital camera (Nikon, USA) connected to an inverted microscope. The image capturing and analysis was performed using an in house software made using Python Programming language

\section*{Appendix B: Data Analysis}

\subsection*{Details about the tail-tracking procedure}

\begin{figure*} 
    \centering
  \includegraphics[width=0.3\linewidth]{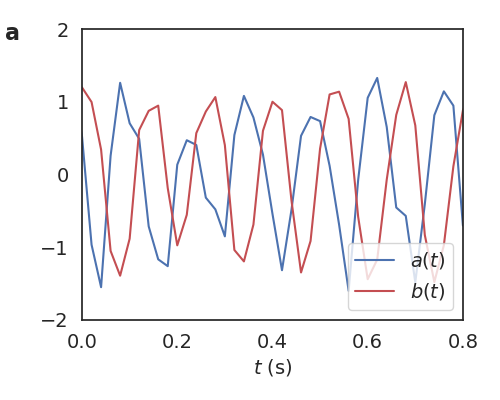}
  \includegraphics[width=0.6\linewidth]{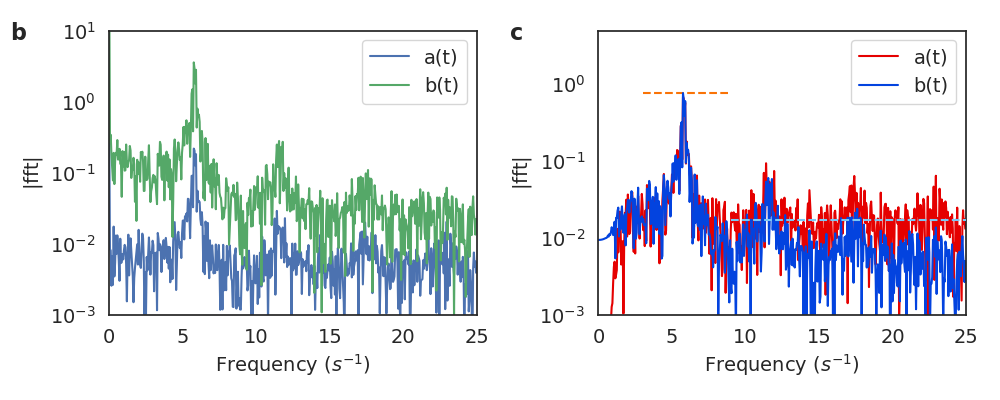}
  \caption{\label{figM3}Details about the tail-tracking procedure. {\bf a}, Examples of the signals $a(t)$ and $b(t)$ from trail tracking: one can see that $a(t)$ anticipates $b(t)$ of an angle between $\sim \pi/2$ and $\sim \pi$.  {\bf b},  spectra (modulus of the fast Fourier transform) of the signals $a(t)$ and $b(t)$. {\bf c},  spectra after signal filtering. The orange and light-blue dashed lines indicate the signal level and the noise level respectively (the noise-to-signal ratio nsr is defined as the ratio between the latter and the former). }
\end{figure*}

Images are collected at 50 frames per second, with 20x objective resulting in a resolution of $6.5/20$ $\mu m$ per pixel. Each image portrays a large portion of the substrate where several cages  are present, almost all filled by caged sperms. Only cages with a single trapped sperm cell are analysed. A region of interest (ROI) of averagely $40\times 40$ pixels - corresponding to an area of roughly $13 \times 13 \mu m ^2$ - containing the most visible part of the tail which is also the one closest to the head (see Fig.~\ref{fig1}d), is cropped and treated by successive layers of image processing tools: 1) background subtraction to reduce noise, 2) transform to gradient (squared modulus)  to avoid dependence on absolute levels, 3) Gaussian filter with $1$ pixel range, 4) the largest continuous bright region is individuated (it always corresponds to the tail), 5) that region is treated as a  cloud of scattered points representing a curve $y$ vs. $x$, which is fitted by least squares to a second order polynomial $y(x,t)=a(t)+b(t)x+c(t)x^2$ as discussed in the main text. The time series of $a(t)$ and $b(t)$ are filtered by a $3$rd order Butterworth high-pass digital filter with critical frequency  set $1.5$ Hz. In Figure~\ref{figM3} the time series of $a(t)$ and $b(t)$ of a tracked cell are shown, together with the  spectra of the two series before and after the filtering. The noise-to-signal ratio (nsr) is computed as the ratio between the noise level and the signal level, both shown in the  Figure.

\subsection*{Discussion of thermal diffusion effects due to the fluid}

The value of $D$ is the result of a complex interplay of elasticity,
hydrodynamics, activity and noises with different origins. Even at
thermal equilibrium, i.e. for dead sperms, an estimate of filament
phase diffusivity is complex as it involves not only the
amplitude of fluctuations, that can be inferred by equilibrium
distribution of elastic energy, but also the relaxation time of such
modes. A first estimate of involved timescales can be obtained by computing the rotational diffusivity. For a passive rod~\cite{elgeti2009self} (or a filament with low flexibility)
of length $\ell \approx 50 \mu m$ (as the sperm's body) in water viscosity
$\eta$ one has a rotational diffusivity of the order $D_{Stokes} \approx k_B T/(\eta \ell^3) \approx 10^{-5}
\textrm{rad}^2/s$.
Visual inspection of our samples   show that
non-motile cells are basically immobile with negligible fluctuations in position or in shape, within our space-time
resolution.

\subsection*{Correlation length based upon the thermodynamic uncertainty relation}

Here we discuss a simple scaling argument to pinpoint the minimal assumptions behind the definition of an uncertainty correlation length 
\begin{equation}
\mathcal{L}_{TUR}=\frac{\dot{W}}{p_N k_B T}  \label{ltur}
\end{equation}
for a system of $N$ connected motors  (e.g. a chain similar to the axoneme structure).
We recall that the asymptotic (steady state) precision $p$  is defined as
\begin{equation}
p_N=\lim_{t \to \infty} \frac{1}{t} \frac{\langle X_N(t) \rangle^2}{\langle X_N^2 \rangle_c}
\end{equation}
where $\langle x^2 \rangle_c$ stands for the variance of variable $x$, and $X_N(t) = \int_0^t ds \dot{X}_N(s)$ is the observed integrated current.

At a first order approximation, the presence of spatial correlations across a correlation length $\tilde{N}$ inside the chain can be accounted for by re-grouping the $N$ motors in $M=N/\tilde{N}$ independent groups. Moreover the observed integrated current $X_N$  can be assumed to be an empirical average of the integrated current  in each of the $M$ independent groups, i.e.
\begin{equation}
X_N=\frac{\sum_{i=1}^{M} X_{\tilde{N},i}}{M},
\end{equation}
with the $X_{\tilde{N},i}$ being independent and identically distributed. 
These assumptions lead to 
\begin{equation}
p_N = \lim_{t \to \infty}  \frac{1}{t} \frac{\langle X_{\tilde{N}}(t) \rangle^2}{\langle X_{\tilde{N}}^2 \rangle_c/M} =  M p_{\tilde{N}}.
\end{equation}
Assuming that the mean consumed work is extensive in the size of the chain, i.e. $\dot{W}_N \sim N \dot{W}_1$, we get
\begin{equation}
\frac{\dot{W}_N}{p_N k_B T}= \tilde{N} \frac{\dot{W}_1}{p_{\tilde{N}}} = (\mathcal{Q}_{micro} )^{-1} \tilde{N}, \label{ltur2}
\end{equation}
having considered
\begin{equation}
\frac{\dot{W}_1}{p_{\tilde{N}}} \approx \frac{\dot{W}_1}{p_1}  = (\mathcal{Q}_{micro} )^{-1}
\end{equation}
In the last passage we have assumed $p_{\tilde{N}}\approx p_1$ following the assumption that for the correlated $\tilde{N}$ motors in a group the precision is that of a single motor. Eq.~\eqref{ltur2} justifies our definition in Eq.~\eqref{ltur}

\subsection*{Previous experiments with sperms at physiological temperature and with other flagella. }

In a recent work~\cite{ma2014active} data from sperm cells observed at 37  °C have been analysed. Such data, collected in a previous work~\cite{riedel2007molecular} concern an anomalous swimming regime which is apparently induced by a particular sample preparation: they were ``incubated with 1\% F-127 (Sigma) in PBS for 5 min.... When the surface was treated with F-127, the sperm did not stick but instead swam close to the surface, usually in circles of radii on the order of 40 $\mu m$''~\cite{riedel2007molecular}. With such treatment the measured beating frequency was particularly high, we denote it  as $f_{Ma} \sim 31$ Hz, much higher than what usually observed (literature reports $20$ Hz for bull sperm cells at 37 °C~\cite{rikmenspoel1984movements}, observed also in~\cite{riedel2007molecular} without such surface treatment). Within such particular conditions the authors measured a quality factor which, in our notation, reads $q=J/2D \sim 38$ that would correspond to a precision $p=2qJ$.


Other experimental works have addressed the properties of noise in the beating of axonemes, particularly with Chlamydomonas flagella~\cite{goldstein2009noise,goldstein2011emergence,wan2014rhythmicity,polin2009chlamydomonas,quaranta2015hydrodynamics} . In~\cite{goldstein2009noise,goldstein2011emergence} the 
quality factor of beating was obtained indirectly from the rate of phase slips in pairs of synchronised flagella (as well as directly from the distribution of beating periods), getting estimates in a range $q=25-120$, with average beating frequency $f=50 Hz$.

A summary of such previous observations and a comparison with the results of the present study is given in Table~\ref{summarytable}. In compiling this table we have used some assumptions typically found in the literature i.e. that the amplitude of sperm's beating is $6 \pm 1 \mu m$, the amplitude ("wingspan") of Chlamydomonas flagellar beating is $10 \pm 1 \mu m$, and the energy consumption in both cases is given by Taylor formula Eq.~\eqref{taylor} multiplied by $10$ (that is assuming an average efficiency of $10\%$).

\begin{table*}
\begin{tabular}{|c|c|c|c|c|c|c|c|}
\hline
Experiment & $f$ & $L$  & $N$ & $\dot{W}/(k_B T)$ & $q$ & $p_{max}$ & $\mathcal{L}_{TUR}$ \\ \hline
Sperms (this study) & $8 \pm 1$ & $60 \pm 3 $ & $1.1 \cdot 10^5$ & $10^7$ & / & $120 \pm 30$ & $8.5 \cdot 10^4$  \\ 
Sperms after $\sim 2$ hours & $6 \pm 1$ & $60 \pm 3 $ & $1.1 \cdot 10^5$ & $5.7 \cdot 10^6$ & / & $60 \pm 10$ & $9.6 \cdot10^4$  \\
Sperm after $\sim 4$ hours & $3 \pm 1$ & $60 \pm 3 $ & $1.1 \cdot 10^5$ & $1.4 \cdot 10^6$ & / & $10 \pm 5$ & $1.4\cdot 10^5$  \\ 
Sperm after $\sim 5$ hours & $2 \pm 1$ & $60 \pm 3 $ & $1.1 \cdot 10^5$ & $6.4 \cdot 10^5$ & / & $4 \pm 0.5$ & $1.6 \cdot 10^5$  \\ \hline
Sperms at 37 °C~\cite{ma2014active} & $31 \pm 1$ & $60 \pm 3 $ & $1.1 \cdot 10^5$ & $1.5 \cdot 10^8$ & $38 \pm 5$ & $1.4 \cdot 10^4$ & $10^4$  \\ \hline
Chlamydomonas ~\cite{goldstein2009noise} & $47 \pm 2$ & $10 \pm 0.5 $ & $1.8 \cdot 10^4$ & $1.6 \cdot 10^8$ & $23 \pm 16$ & $1.3 \cdot 10^4$ & $1.2\cdot 10^4$  \\ 
Chlamydomonas ~\cite{goldstein2009noise}  & $47 \pm 2$ & $10 \pm 0.5 $ & $1.8 \cdot 10^4$ & $1.6 \cdot 10^8$ & $26 \pm 18$ & $1.6 \cdot 10^4$  & $10^4$ \\ \hline
Chlamydomonas ~\cite{goldstein2011emergence} & $71 \pm 2$ & $6 \pm 0.5 $ & $10^4$ & $2.3 \cdot 10^8$ & $70 \pm 10$ & $6.2 \cdot 10^4$ & $3.6\cdot 10^3$  \\ 
Chlamydomonas ~\cite{goldstein2011emergence}  & $67 \pm 2$ & $8 \pm 0.5 $ & $1.4 \cdot 10^4$ & $2.6 \cdot 10^8$ & $100 \pm 12$ & $8.3 \cdot 10^4$  & $3.1 \cdot10^3$ \\ 
Chlamydomonas ~\cite{goldstein2011emergence}  & $62 \pm 2$ & $10 \pm 0.5 $ & $1.8 \cdot 10^4$ & $2.9 \cdot 10^8$ & $120 \pm 18$ & $9.6 \cdot 10^4$  & $3  \cdot10^3$ \\ \hline
Chlamydomonas ~\cite{wan2014rhythmicity}  & $60 \pm 3$ & $10 \pm 0.5 $ & $1.8 \cdot 10^4$ & $2.7 \cdot 10^8$ & $100 \pm 20$ & $7.5 \cdot 10^4$  & $3.5  \cdot10^3$ \\ \hline
Chlamydomonas ~\cite{quaranta2015hydrodynamics}  & $53 \pm 2$ & $10 \pm 0.5 $ & $1.8 \cdot 10^4$ & $2 \cdot 10^8$ & $199 \pm 20$ & $1.3 \cdot 10^5$  & $1.6  \cdot10^3$ \\ \hline
\end{tabular}
\caption{ \label{summarytable} Summary table of results from previous literature and from the present studies for both sperms and Chlamydomonas. The columns report the kind of experiment (with reference), the frequency of beating (in Hz), the length of the flagellum (in $\mu m$), the estimated number of motors $N=1800 L$, the energy consumption in units of $k_B T$, the quality factor $q$ (when measured), the maximum observed precision $p_{max}$ (in $s^{-1}$) and the precision-based correlation length $\mathcal{L}_{TUR}=\dot{W}/(p k_B T)$ in number of motors, see Appendix B.3. For reasons of space we have not reported the errors if they can be computed by standard error propagation. The reported errors are those obtained from estimate reported in the literature and standard deviations in our experimental observations.}
\end{table*}

A plot of the TUR-based correlation length versus the length of the flagellum is shown in Figure~\ref{figM1}. Within the error, the data are compatible with long-range order, i.e. $\mathcal{L}_{TUR} \sim N$.

\subsection*{Estimates of the energy consumption and efficiency of sperm swimming. }

Consumption rate, speed and beating frequency are sensitive to environmental conditions, e.g. temperature and
fluid viscosity~\cite{rikmenspoel1984movements}. In~\cite{brokaw1967adenosine} sea urchin sperm was studied in a 50\% glycerol solution at $16^\circ C$ with varying the beating frequencies through modulation of the ATP concentration: for instance at $20$ Hz it was found $10^6$ molecules of ATP per sperm per second, corresponding to slightly more than  $\approx 10^7 k_B T$ per second. In~\cite{rikmenspoel1969energy} experiments were performed at $37^\circ C$, with bull semen diluted/washed in egg yolk with diluents, a phosphate buffer and the addition of fructose and lactate, leading to an estimate of consumption rate equal to $\approx 10^7$ molecules of ATP per sperm per second, i.e. slightly more than $\approx 10^8 k_B T$ per second. The evaluation of the produced work through the Taylor formula led to an estimate of the efficiency of $\sim 20\%$.  In~\cite{chen2015atp} sea urchin  sperms are studied one by one in droplet solutions, at unreported temperature but with controlled conditions in both ATP concentration and buffer viscosity (both directly modulating the beating frequency), obtaining $\sim 3 \cdot 10^6$ ATP molecules per sperm per second when the tail beats at  $10$ Hz.

\section*{Appendix C: A theoretical model for the fluctuations of an active axoneme}

The model discussed in this section is a variation of  the classical model introduced in~\cite{julicher1997spontaneous} and further studied in~\cite{guerin2011dynamical} and~\cite{guerin2011bidirectional}.

\begin{figure*}  
    \centering
  \includegraphics[width=0.33\linewidth]{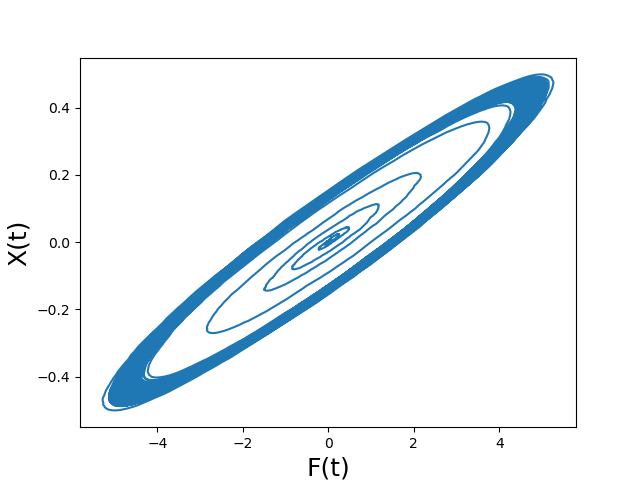}
  \includegraphics[width=0.3\linewidth]{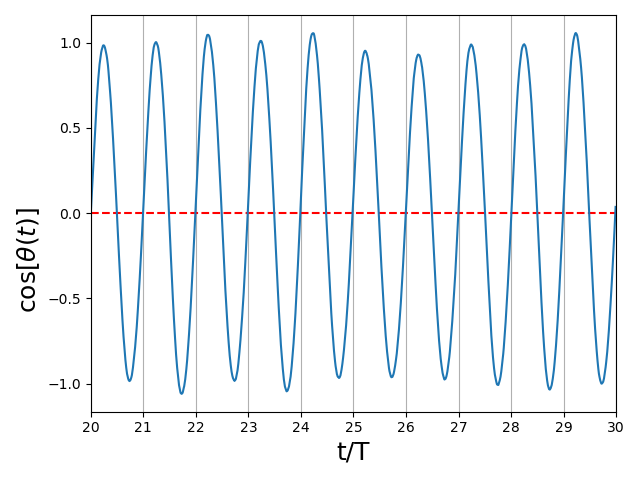}
  \includegraphics[width=0.33\linewidth]{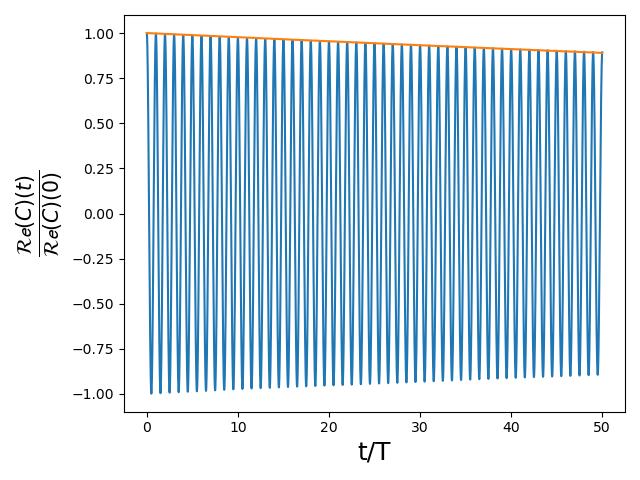}\\
   \includegraphics[width=0.33\linewidth]{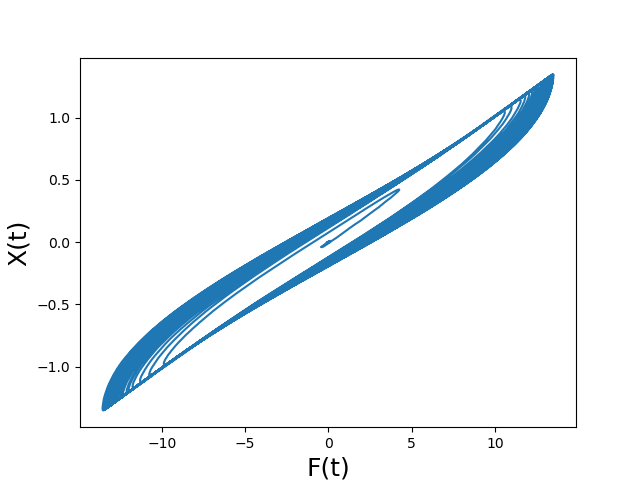}
  \includegraphics[width=0.3\linewidth]{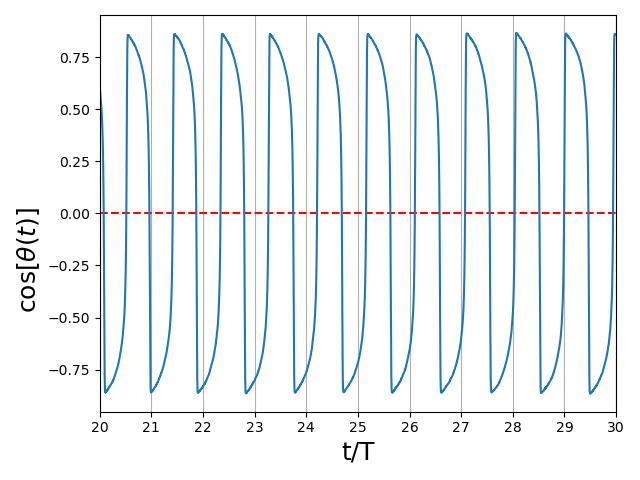}
  \includegraphics[width=0.33\linewidth]{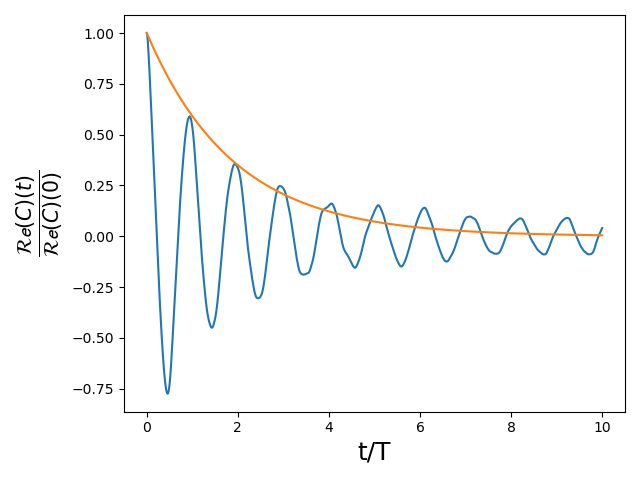}
    \caption{\label{figM2}Theoretical model. {\bf a,d}, evolution of the system in the force-coordinate phase space ($K=0$ in a and $K=10$ in d), the angle $\theta(t)$ measures the phase of this limit cycle after suitable rotation and normalisation of axis;  {\bf b,e}, evolution of $\cos[\theta(t)]$ that allows to evaluate the  stability of the periods in the two cases  ($K=0$ in b and $K=10$ in e);  {\bf c,f}: real part of the autocorrelation of $e^{i \theta(t)}$ and exponential fit $\sim e^{-D t}$ of its envelope ($K=0$ in c and $K=10$ in f).  In all the simulations we have used $\alpha=\eta=0.5$, $k/(\xi \Omega)=10$, $\alpha N U_0/(\Omega \ell^2 \xi)=0.6$. }
\end{figure*}

Interestingly the original model has been used to rationalise recent experiments on sperm swimming fluctuations~\cite{ma2014active}. In the original model however, the fluctuations of the $N$ motors are independent, therefore the fluctuations of the filament macroscopic dynamics are somehow similar to the fluctuations of an {\em average} of  $N$ independent noises, therefore their squared error  (or diffusivity) decreases with $N$ and this result in a linear $\sim N$ increase of the precision (or quality factor).  We provide a simple mechanism to couple the noises of the motors and verify, in numerical simulations, that this ingredient is sufficient - at strong coupling - to change the  size scaling from $O(N)$ to $O(1)$.

In the model the filament is represented by a position $X(t)$ and by a potential $W=\sum_{i=1}^N s_i U[x_i-X(t)]$ which regulates the interaction of the filament with $N$ motors, each one being at fixed position $x_i$ and in attachment state $s_i=0,1$  ($0$ when detached and $1$ when attached). The position $X(t)$ can be understood as the real position in space of  the center of mass of the  filament, as well as a generalised coordinate representing the shape of it. The potential may be related to  local properties of the filament, such as the local curvature which depends upon the time  $t$ through the coordinate $X(t)$. Each motor can detach and re-attach from/to the filament, changing its state $s_i$, according to a Poisson process that violates detailed balance, as it happens when ATP $\to$  ADP+P  process is involved. The overdamped equation  of motion of the filament is $\xi\dot X(t)=F_{ext}(t)+F(t)$ where  $F(t)=-\partial_X \sum_i s_i U[x_i-X(t)]$ and $F_{ext}(t)$ is an external force. In general the filament can be free from external forces, but then a spatial asymmetry (employed in $W(x)$) is needed to induce forward motion, otherwise an external force (e.g. a spring on an end of the filament) is already sufficient to break spatial symmetry and the potential $W$ can be taken symmetric to simplify calculations. This is the case analysed here and in~\cite{ma2014active}, with $F_{ext}=-\kappa X$ and $U(x)=U_0[1-\cos(2\pi x/\ell)]$, the filament  does not move  on average but fluctuates more or less regularly, while a limit cycle in the plane $X(t),F(t)$ can be used as analogous to the $A,B$ plane used in our experiments. In the original model, each motor realises the attachment/detachment process independently from the other motors, with the only indirect correlations due to the modulation of the attachment/detachment rate through the position $\omega_{on}^i=\Omega[\eta -\alpha \cos(2\pi (x_i-X(t))/\ell)]$ and $\omega_{off}^i=\Omega-\omega_{on}^i$, . This ingredient however only correlates (locally) the average residence times but does not correlate fluctuations around those averages: it is the same as considering independent noises with similar averages.

In order to adapt the model to our experimental findings, we introduced  a binding potential that correlates adjacent motors: this potential is minimised when  adjacent  motors are in  the same state. This is implemented as a modification of the rates according to the  formula  $\omega_{on}^i=\Omega[\eta -\alpha \cos(2\pi (x_i-X(t))/\ell)]e^{-\Delta U^{bind}_i}$ and $\omega_{off}^i=\Omega-\Omega[\eta -\alpha \cos(2\pi (x_i-X(t))/\ell)]e^{-\Delta U^{bind}_i}$, with  $\Delta U^{bind}_i$ is the binding potential increase after the variation of state $s_i$ of the  $i$-th motor, and the binding potential is $U^{bind}_i=K(s_i-s_{i+1})^2+K(s_i-s_{i-1})^2$. When $K=0$ the original model without binding energy is recovered. 

The effect of $K$ can be appreciated in numerical simulations of the model whose results are reported in Fig.~\ref{fig4} and~\ref{figM2}. In particular in Figure~\ref{figM2}c-f we show the drastic change in the decay of the phase correlation when $K$ is increased. In Fig.~\ref{fig4}b it can be appreciated how the size scaling of the precision changes completely and tends to become {\em independent} of $N$ when the coupling strength increases. Our observation that the macroscopic sperm precision ($N\sim 10^5$) is similar to the microscopic sperm precision ($N=1$) is fairly explained by this new model. Note that the beating frequency in the model is independent of $N$ (at least for $N \gtrsim 10^2$) so that the energy consumption (ATP consumed per cycle and per motor) increases with $N$, even for large binding potential. The ratio between precision $p \sim O(1)$ and energy consumption  $\dot W \sim N$ is therefore doomed to decrease as $1/N$, in fair agreement with experimental observations.

\section*{Appendix D: The simplest working principle for  Thermodynamic Uncertainty Relations}

While the TURs have been demonstrated for larger and larger classes of models and time domains, we judge instructive to summarise the first example where they have been observed which is a Markov jump process describing - in a very simplified way - the stochastic (progressive  on average) dynamics of a single Brownian motor or clock~\cite{barato2015thermodynamic}. The model is defined in continuous time, the motor  can go forward or backward with probability rates $k_+$ and $k_-$ respectively. Local detailed balance dictates $k_+/k_- \sim \exp(Q/k_B T)$ where $Q=W$ is the energy dissipated in a forward jump equal to the work input carried by ATP. 

The average current of the clock (number of steps per unit of time) is $J=k_+-k_-$, while the associated diffusivity is $2D=k_+ + k_-$, therefore for the position $X(t)$ of the motor/clock one has a relative uncertainty defined as $\epsilon^2=(\langle X^2\rangle-\langle X \rangle^2)^2/\langle X \rangle^2$ which reads $\epsilon^2 = 2Dt/(Jt)^2=(k_+ + k_-)/[(k_+-k_-)^2t]$ which in terms of the precision $p=2/(t \epsilon^2)$ reads $p=2(k_+-k_-)^2/(k_++k_-)$. The energy dissipated up to time $t$ reads, in terms of the entropy production rate $\sigma$, $T\sigma t=\dot Q t = JQ t$. Then the product between the energy dissipated and the relative uncertainty satisfies $2 \dot Q/p=T\sigma t \epsilon^2 =  (k_++k_-)/(k_+-k_-) Q \coth[Q/(2 k_B T)] \ge 2 k_B  T$ which leads to the TUR used in this paper $p \le \dot Q/(k_B T)$.

This example is useful to evaluate the key sources of noise in this process, i.e. the contributions to $D$ which are both $k_+$ and $k_-$. This means that backstepping (a not negligible $k_-$) is not  the only source of noise, but  $k_+$ also contributes to noise. The reason is that a large contribution to fluctuations of the motor current is due to fluctuations in the residence time before a new forward step. If time is discretized in $dt$ steps, the motor remains in its position with a probability $1-(k_+)dt-(k_-)dt$: the exit time has an exponential probability with average exit rate $(k_+)+(k_-)$. The real chemical network of a molecular motor, such as the dynein, is much richer than the minimal model considered in~\cite{barato2015thermodynamic,howard2001,peliti2021stochastic}: in that minimal model a single step is a coarse-graining of the several intermediate chemical steps. The presence of intermediate steps with their fluctuating residence of times and possibly non-negligible backstepping probabilities implies relevant fluctuations in the coarse-grained residence times and therefore in the motor's current, even when the total back-stepping probability is negligible. We deem these motors' fluctuations, with an additional coordination hypothesis discussed in the text, to be important for the deviation of the sperm shape cycle from its average dynamics.


\newpage

\newpage

\section*{SUPPLEMENTAL MATERIAL}

\subsection*{Tracking the head oscillations}

In a series of experiments we have tracked the sperm’s head instead of the tail, finding  results in qualitative and quantitative agreement with those reported in the main article. The setup is identical, but different lighting and focus conditions with the microscope allowed  us to have more image contrast in the interior of the  cages. Images are collected with the same objective and ccd as in the main experiment. The region of interest is averagely the  same area as in the main experiment. Background subtraction reduces the visibility of cage’s boundaries. The covariance matrix of the pixel light distribution is computed: its eigenvector associated to the maximum eigenvalue determines the direction of orientation of the sperm’s head, whose slope replaces coefficient $-b(t)$ of our tail’s analysis (the minus sign of course depends upon the reference frame we are using). The $\hat{y}$ position of the center of mass of the head replaces coefficient $a(t)$. Once we get $a(t),b(t)$ for the image at time $t$  we can repeat exactly the same analysis as we did for the tail’s tracking. In Figure S1 we summarise the results of this analysis.

\begin{figure*}
\includegraphics[width=16cm]{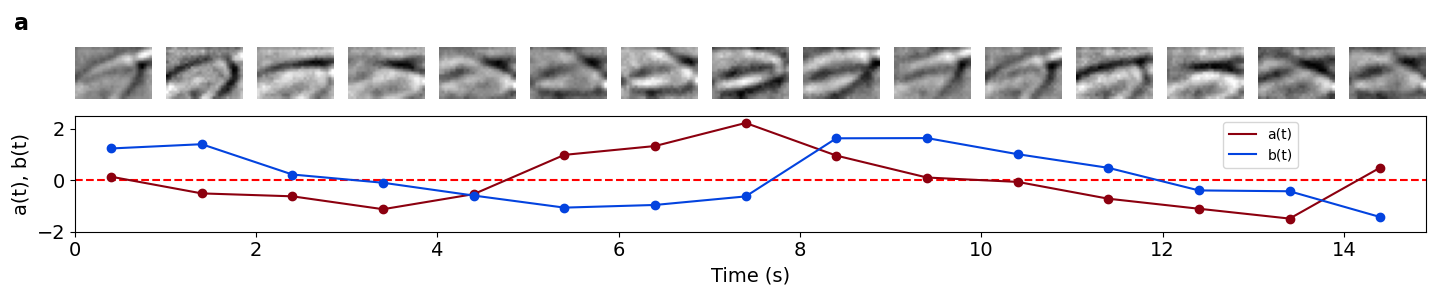} 
\includegraphics[width=10.5cm]{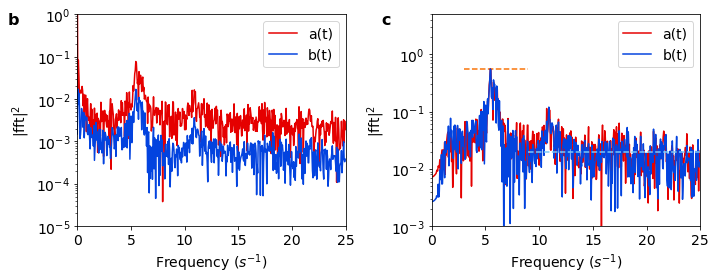} 
\includegraphics[width=5cm]{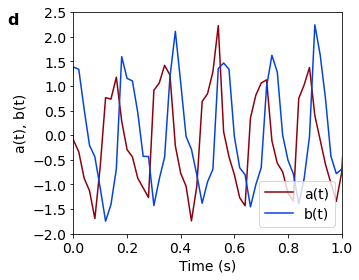}
\includegraphics[width=5cm]{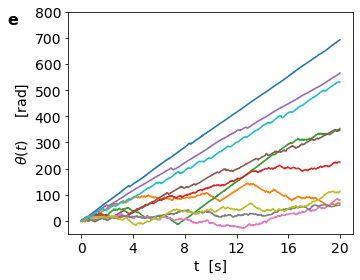}
\includegraphics[width=5cm]{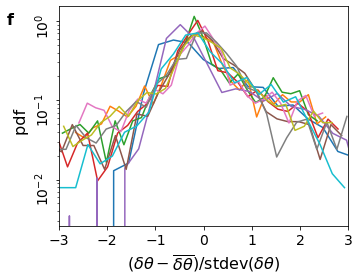}
\includegraphics[width=5cm]{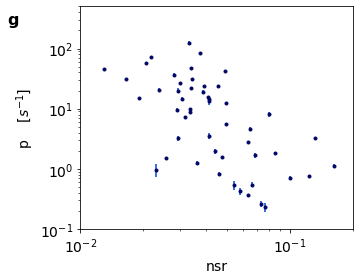}\\
\noindent Fig. S1 Head tracking.  {\bf a},  Example of a head tracking where $a(t)$ and $b(t)$ (bottom plot) are obtained by computing the height ($y$ position in the region of interest) of the center of mass of the pixel distribution and the slope of the direction associated with the maximum eigenvalue of the pixel covariance matrix. {\bf b}, spectra of the signals $a(t)$ and $b(t)$. {\bf c}, spectra after signal filtering. The orange and light-blue dashed lines indicate the signal level and the noise level respectively. {\bf d}, Examples of the signals $a(t)$ and $b(t)$ from head tracking: again $a(t)$ anticipates $b(t)$ of an angle between $\sim \pi$ and $\sim \pi/2$. {\bf e}, Integrated phase-space current $\theta(t)$ for a few
  observed sperms. {\bf f}, Pdf
  (over $20$ seconds acquisition), for a few sperms, of $\delta
  \theta=\theta(t+dt)-\theta(t)$ (shifted by the mean and scaled by
  the standard deviation), where $dt=0.02$ seconds. {\bf g},  Precision
  $p$ versus the noise-to-signal ratio (nsr) computed from the signals
  $a(t)$ and $b(t)$. The sample size consists of $58$ different observed sperm cells.
\end{figure*}

\subsection*{Alternative estimates of the precision}

In a recent paper the fluctuations of the sperm beating cycle have been analysed~\cite{ma2014active}. The Authors suggest two alternative ways to estimate the phase diffusivity $D$ which is necessary to get values for $p=J^2/D$ (being $J=2\pi f$ and $f$ the average beating frequency). A first, perhaps more approximate, procedure is to retrieve the quality factor $q=f_0/\Delta f_0$ where $\Delta f_0$ is the width-at-half-maximum in the power spectrum, which reads $\Delta f_0 \sim 3$ Hz giving $q  \approx 3$. The quality factor is related to the phase diffusivity by $q=J/(2D)$ which leads to $p=J^2/D=2Jq \approx 2 \times 10^2 s^{-1}$  which is compatible in order of magnitude with the maximum precision we report in the main text. The second recipe consists in measuring the decay in time of the phase correlation $C(t)=\langle \exp i [\theta(t_0+t)-\theta(t_0)] \rangle \sim e^{-D t}$, as we have also done for the simulations of the theoretical model in the Main text. Examples of the decay of $C(t)$ are shown in  Figure S2. The obtained values of $D$ lead to a precision of the order of $p_{max} \sim 5 \times 10^2 s^{-1}$ and smaller by a factor  $\sim 10$ after $5$ hours of oxygen deprivation, in full agreement with the results shown in the Main text.

\begin{figure*}
\includegraphics[width=5.2cm]{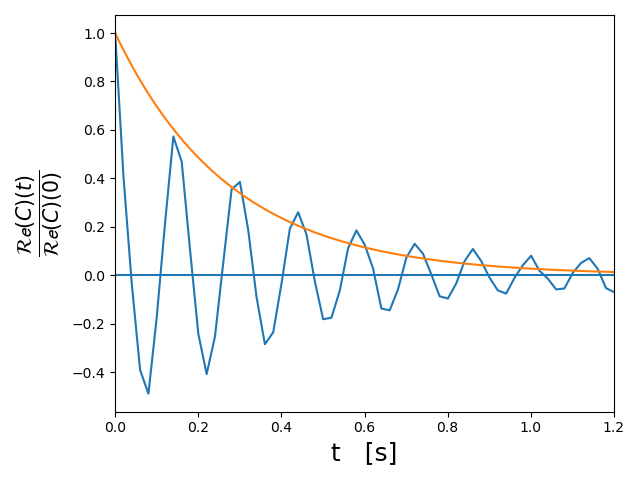}
\includegraphics[width=5.2cm]{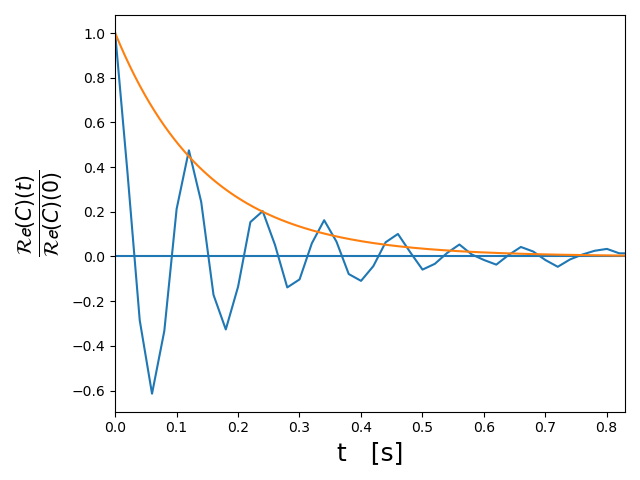}
\includegraphics[width=5.2cm]{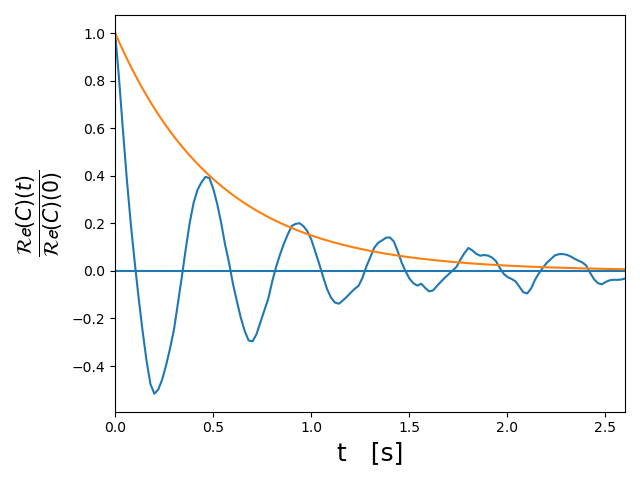}

\noindent Fig. S2 Phase correlation decay for three different sperms in experiments. The first two are at the starting time of the experiment, while the third is after $150$ minutes. The precisions measured are $\sim 500 s^{-1}$, $\sim 400 s^{-1}$ and $\sim 80 s^{-1}$. 

\end{figure*}

\begin{table*}
\begin{tabular}{|l|l|}
\hline
$\theta$ & integrated current (beating phase) \\
$J$ & average current \\
${\textrm msd}$ & mean squared displacement of $\theta$ \\
$D$ & diffusivity of $\theta$ \\
$p$ & precision \\
$p_{max}$ & estimate of the largest precision in the sperm's population \\
$p_{TUR}$ & theoretical bound for the precision, based upon the TUR \\
$\dot W$ & energy consumption rate \\
$T$ & temperature \\
$k_B$ & Boltzmann's constant \\
$\textrm{nsr}$ & noise to signal ratio \\
$\eta$ & fluid viscosity \\
$L$ & length of the flagellum \\
$f$ & beating frequency \\
$\beta$ & beating amplitude \\
$\mathcal{Q}$ & TUR-based figure of merit, i.e. $p_{max}/p_{TUR}$ \\
$\mathcal{Q}_{macro}$ & TUR-based figure of merit for the whole flagellum or large chain of motors\\
$\mathcal{Q}_{micro}$ & TUR-based figure of merit for the single motor\\ 
$X$ & relative position between motors and flagellum in the theoretical model\\
$\xi$ & viscous damping in the theoretical model \\
$\kappa$ & elastic constant in the theoretical model \\
$s_i$ & state variable ($0$ detached, $1$ attached) for the $i$-th motor in the theoretical model \\
$U(x)$ & potential energy in the theoretical model \\
$N$ & number of motors in the experiments (estimate) or in the theoretical model \\
$K$ & coupling constant for the binding potential that couples the state of two adjacent motors \\
$\mathcal{L}_{TUR}$ & putative correlation length defined according to the TUR \\
$k_+,k_-$ & probability rates for transitions in the simple motor model in Appendix D \\
$\epsilon$ & relative uncertainty in the simple motor model in Appendix D \\
$\sigma$ & entropy production rate in the simple motor model in Appendix D \\
$Q$ & dissipated heat per cycle (equal to work input per cycle) in  Appendix D \\
$q$ & quality factor  $q=J/(2D)=p/(2J)$ \\
\hline 
\end{tabular}
\caption{List of symbols used in the text}
\end{table*}

\bibliography{spermprecision}


\end{document}